\documentclass[11pt]{article}
\usepackage{epsfig,amssymb,amsmath,psfrag,float,etoolbox,amsfonts,amsbsy,bm,euscript,mathrsfs,stmaryrd,faktor,graphics,tikz}

\definecolor{hyperref}{RGB}{026,028,185}
\usepackage[bookmarks=true,colorlinks=true,linkcolor=hyperref,citecolor=hyperref,urlcolor=hyperref,bookmarksnumbered]{hyperref}
 
\def\ga{\gamma}

\newcommand{\cN}{{\cal N}}

\def\theequation{\thesection.\arabic{equation}}
\def \be  {\begin{equation}}
\def \ee  {\end{equation}}
\def \ba  {\begin{eqnarray}}
\def \ea  {\end{eqnarray}}
\def\bea{\begin{eqnarray}}
\def\eea{\end{eqnarray}}

\def \ln {{\rm ln}}

\def\ln{ \log }

\def\as{\alpha_s}
\def\ifm{\ifmmode}
\def\msb{\ifm \overline{\rm MS}\, \else $\overline{\rm MS}\,$\fi}

\def\g{\gamma}
\def\m{\mu}

\def\la{\lambda}
\def\N{\mathcal{N}}

\begin{document}

\thispagestyle{empty}

\vspace{ -3cm} \thispagestyle{empty} \vspace{-1cm}
\begin{flushright} 
\footnotesize
HU-EP-13/19\\
\end{flushright}%

\begingroup\centering
{\Large\bfseries\mathversion{bold}
On DIS Wilson coefficients 
in $\cN=4$ super Yang-Mills theory
\par}%
\vspace{7mm}

\begingroup
Lorenzo~Bianchi$^{a}$\footnote{lorenzo.bianchi@physik.hu-berlin.de}, Valentina~Forini$^{a}$\footnote{valentina.forini@physik.hu-berlin.de}, Anatoly V.~Kotikov$^{b}$\footnote{kotikov@theor.jinr.ru, kotikov@mail.desy.de}\\
\endgroup
\vspace{8mm}
\begingroup\small
$^{a}$ \emph{Institut f\"ur Physik, 
Humboldt-Universit\"at
zu Berlin\\ Newtonstra{\ss}e 15, 12489 Berlin, Germany}\\
$^{b}$ \emph{Bogoliubov Laboratory of Theoretical Physics, 
Joint Institute for Nuclear Research\\
141980 Dubna, Russia}
\endgroup


\textbf{Abstract}\vspace{5mm}\par
\begin{minipage}{14.7cm}

In this note we evaluate Wilson coefficients for ``deep inelastic scattering'' (DIS) in $\mathcal{N}=4$ SYM theory at NLO in perturbation theory, using as a probe an R-symmetry conserved current. They exhibit uniform transcendentality and coincide with the piece of highest transcendentality in the corresponding QCD Wilson coefficients.  We extract from the QCD result a NNLO prediction for the $\mathcal{N}=4$ SYM Wilson coefficient, and comment on the features of its Regge limit asymptotics.

\end{minipage}\par
\endgroup


\section{Discussion}
\setcounter {equation} {0}

Among the many ways through which the outstanding simplicity of the $\N=4$ super Yang-Mills (SYM) theory reveals itself,  the pattern of transcendentality exhibited by many of the observables computable in a close analytical form has the merit of setting potentially a quite direct link to QCD. In particular, the maximum transcendentality principle~\cite{Kotikov:2002ab,Kotikov:2004er} (MTP) is the conjecture -- inspired by special properties~\cite{Kotikov:2000pm} for the maximally supersymmetric generalization of BFKL and evolution equations -- that in the anomalous dimensions of leading twist operators only terms of highest transcendentality arise, which can  be picked up by the `most complicated' terms of the corresponding QCD results~\cite{Moch:2004pa,Vogt:2004mw} with the appropriate color factor prescription $C_A=C_F=N_c$ and $T_f\,n_f=2N_c$. 
Here a transcendentality weight $n$ is given to each Riemann $\zeta$ value $\zeta_n\equiv\zeta(n)$, with a similar tallying for the harmonic sum $S_{\vec{n}}(j)$,  and the principle states that the anomalous dimension $\gamma(j)$ at $n$ loops is a linear combination of harmonic sums of transcendentality $2n-1$.
Several signals of consistency~\cite{Staudacher:2004tk,Bern:2005iz,Eden:2006rx, Beisert:2006ez, Kotikov:2006ts,Bern:2006ew} have lead to assign a predictive power to the MTP, which (combined with other QCD-related properties~\cite{reciprocity}) has long been the computational strategy for extracting multi-loop anomalous dimensions of twist operators from algebraic Bethe equations~\cite{Kotikov:2007cy,Bajnok:2008qj,Beccaria:2009eq,Lukowski:2009ce}.\\
 Remarkably, patterns of uniform leading transcendentality  -- in this case the degree of transcendentality is $2n$ for $n$-loop results -- appear also in $\N=4$ SYM scattering amplitudes~\cite{Bern:2006ew,Dixon:2011xs}~\footnote{See also~\cite{Dixon:2009ur} for an example of how transcendentality imposes strict limitations on the admissible functional forms of the soft anomalous dimension.} even at the subleading-color (non-planar) level~\cite{Naculich:2008ys}, in light-like Wilson loops~\cite{Drummond:2007cf}~\footnote{See also \cite{Bianchi:2013pva}.} as well as in form factors~\cite{Bork:2010wf, Gehrmann:2011xn, Brandhuber:2012vm} and correlation functions~\cite{Eden:2012rr,Drummond:2013nda}~\footnote{See also the field-theory related structure observed for superstring amplitudes~\cite{Schlotterer:2012ny}.}. \\
Significantly, the as yet heuristic nature of the observations and the presence of some exception to the rule of ``direct extraction'' from QCD results~\footnote{In the case of scattering amplitudes, see the mismatch~\cite{DixonPrivate}, mentioned in~\cite{Naculich:2008ys},  between the terms with highest transcendentality  in the leading-color two-loop QCD amplitude of~\cite{Bern:2002tk} and the corresponding $\N=4$ SYM amplitude.} make the detection of such generalized MTP an always necessary, preliminary check to set its predictive power in the context at hand.  

We have here evaluated the simplest non-trivial (NLO) contribution to ``deep inelastic scattering'' Wilson coefficients  
for $\N=4$ SYM, using as external hard probe the current conserved under the internal $SU(4)$ R-symmetry of the theory.   Although in a conformal theory bound-state ``hadrons'' do not form, one can imagine to redefine the unphysical asymptotic states~\footnote{See also~\cite{Bork:2009nc}.} via quark/gluon distributions, governed by the $\mathcal{N}=4$ SYM DGLAP equations~\cite{Kotikov:2002ab}, in which all emerging collinear divergencies are factorized out~\footnote{ In an operative sense, this logic is not different from what one experiences in QCD, where the true, physical, asymptotic states are not reachable via perturbation theory and are actually defined via  partonic distribution functions.}. Looking at the final result (\ref{result1})-(\ref{result3}), of immediate evidence are a) its extreme simplicity (all rational terms in $z$ cancel and only logarithms of $z$ and $1-z$ survive), b) its uniform transcendentality-degree two, evident in its Mellin transform (\ref{resmel1})-(\ref {resmel3}), and c) the fact that it coincides with the highest transcendental part of the QCD result (\ref{CqQCD})-(\ref{CgQCD}), provided the appropriate set of splitting functions reflecting the field content of the theories is taken into account and with the color factor prescription mentioned above.  These are not a priori obvious observations. At least in the case of anomalous dimensions for leading(two)-twist operators -- which govern the IR-divergent $O(1/\epsilon)$~\footnote{We use the FDH dimensional reduction scheme of~\cite{Bern:1991aq,Bern:2002zk}.} part of the inclusive DIS cross-section --
 the maximal transcendentality property seems to be deeply connected to the maximal supersymmetry of the gauge theory (MTP is broken already in the $\mathcal{N}=2$ case~\footnote{Private communication by L.~Lipatov.}). While 
in this sense the $1/2$ BPS~\cite{Basu:2004nt}  current (\ref{current}) is the best candidate to gain  properties a) and b) for the  $O(1/\epsilon)$ contribution~\footnote{It is not difficult to check that using the  R-symmetry singlet, \emph{non conserved}, current $\bar\psi\,\gamma^\mu \psi$ would add to  the formulas obtained, together with a UV divergent part due to bubble diagrams, also terms of transcendentality one and zero.} it is non-trivial that they extend to the IR-finite part.  It is also non-obvious that the maximal transcendentality observed for $\mathcal{N}=4$ SYM form factors~\cite{Gehrmann:2011xn}  would be maintained in the cross-sections here analyzed.
Interestingly, using the same basis (in Mellin space) which diagonalizes the twist-two anomalous dimensions of $\mathcal{N}=4$ SYM ~\cite{Kotikov:2002ab}, the result simplifies further and can be written -- in total analogy with the anomalous dimension case -- in terms of a \emph{universal}  function $C_{uni}(j)$ expressed as combination of harmonic sums and appearing with shifted arguments in the first two entries of the Wilson coefficient vector \eqref{resrot}. Even more interesting are the vanishing of the third vector component, and a shift in the first entry which cancels the singular behavior in the Regge limit. Elaborating on these features and extrapolating them to the next order in perturbation theory we propose~--~extracting it from the QCD result~\cite{Moch:1999eb}~--~the DIS Wilson coefficient for $\mathcal{N}=4$ SYM at NNLO, verifying its self-consistency  
with the analysis of its Regge ($j\to1$) and quasi-elastic ($j\to\infty$) asymptotics .  
 
Since it usually requires two non-trivial orders to reliably identify the general structure of a perturbative computation,  it would be important to  extend the calculations to NNLO and confirm the predictions \eqref{Cuni2} and \eqref{+2a}, and with them the properties described above.  Furthermore, one could proceed generalizing this computation to  crossing-related processes such as the supersymmetric generalization of the $e^+ e^-$ annihilation and Drell-Yan lepton-pair production~\cite{inprogress}.

This note proceeds with the presentation of the main result in $\mathcal{N}=4$ SYM, its comparison to QCD (Section \ref{sec:results}) and the extraction from QCD of an NNLO prediction (Section \ref{sec:diagonalization}).
Appendices A, B and C recall, respectively, the LO expressions for splitting functions appearing in our analysis, a convenient basis for expressing our result and our prediction and basic definitions of harmonic sums.


\section{DIS Wilson coefficients in $\N=4$ SYM at NLO}
\label{sec:results}
\setcounter {equation} {0}

In the QCD deep inelastic scattering analysis, the Wilson coefficients $C_q, ~C_g$ are the short-distance functions which appear  in the IR-safe part of the structure functions, where they multiply respectively the (renormalized) quark and gluon distributions.  At order ${\cal O}(\as)$ and in \msb factorization scheme, the structure function $F_2$  reads
\be
\!\!\!\!
 F_2(x,Q^2)=x\sum_{q,\bar q}\,e^2_q\int_x^1\frac{dz}{z}\,q\Big(\frac{x}{z},\mu^2\Big)\hat{F}_q\Big(\frac{Q^2}{\mu^2},z\Big)+x\sum_{q,\bar q}\,e^2_q\int_x^1\frac{dz}{z}\,g\Big(\frac{x}{z},\mu^2\Big)\hat{F}_g\Big(\frac{Q^2}{\mu^2},z\Big) \label{F2QCD}
\ee
where $x$ is the Bjorken variable $x=Q^2/(2p\cdot q)$ and
\begin{align}
\hat{F}_q\Big(\frac{Q^2}{\mu^2},z\Big)&=\delta(1-z)+ \frac{\as}{4\pi}\,\Big(P_{qq}^{(0)}(z)\ln\frac{Q^2}{\m^2}+C_q^{QCD} (z)\Big)+\mathcal{O}(\as^2), \label{Fq}\\
\hat{F}_g\Big(\frac{Q^2}{\mu^2},z\Big)&=\frac{\as}{4\pi}\,\Big(P_{qg}^{(0)}(z)\ln\frac{Q^2}{\m^2}+C_g^{\rm QCD} (z)\Big)+\mathcal{O}(\as^2).\label{Fg}
\end{align}
Here $\mu$ is the factorization scale~\cite{Collins:1984kg}, the LO non-singlet splitting functions $P_{qq}^{(0)}, P_{qg}^{(0)}$ are reported in Appendix \ref{app:splittingfunctions} and $C_q, ~C_g$  are given by~\cite{Bardeen:1978yd,KubarAndre:1978uy} 
\ba\label{CqQCD}
 C_q^{\rm QCD} (z)&=&C_F \Big[-2 \,p_{qq}(z)\,\Big(\ln \big(\frac{z}{1-z}\big)+\frac{3}{4}\Big)+\frac{9}{2}+\frac{5}{2}z
 -(9+4\,\zeta_2)\,\delta(1-z)\Big]~,\\\label{CgQCD}
 C_g^{\rm QCD} (z)&=&T_R \Big[-2\,p_{qg}(z)\,\Big(\ln\big(\frac{z}{1-z}\big)+4\Big)+6\Big]~.
\ea
Above, $p_{qq}, p_{qg}$ are the polinomials (\ref{psmallqq})-(\ref{psmallgg}) usually introduced in literature and representing the highest transcendental part of the splitting functions.
As a result of factorization, both splitting functions and Wilson coefficients $C_q,C_g$ can be evaluated perturbatively squaring and integrating over the phase space the relevant form factors. At NLO, the processes contributing to $C_q$ are of real gluon emission $\gamma^*+q\to q+g$ and virtual gluon correction $\gamma^*+q\to q$, while the initial gluon process $g+\gamma^*\to q+\bar q$ contributes to $C_g$ (see for example~\cite{Ellis_QCD}).

To mimic the electromagnetic interaction in the $\N=4$ SYM theory case~\cite{CaronHuot:2006te,Bartels:2008zy} we will  substitute the electromagnetic current with the $SU(4)_R$ conserved current
\begin{equation}\label{current}
j^{\mu,I}=\bar{\psi} \,T^I\, \gamma^\mu \psi-\frac{1}{2}\,\phi\,T^I\,(-i\, \overleftarrow{D}^\mu +i\,\overrightarrow{D}^\mu)\phi, \qquad\qquad I=1,...,15~,
\end{equation}
where a summation over the $SU(N_c)$ and $SU(4)_R$ indices is understood.
Notice that, although with the choice above we are intentionally \emph{not} selecting a $U(1)$ subgroup of the internal $SU(4)_R$ symmetry so that all the fermions and scalars of the theory are ``charged'' on equal foot~\footnote{In~\cite{CaronHuot:2006te}, the U(1) subgroup generated by the diagonal generator $t_3\equiv{\rm diag}(1,-1,0,0)$ was selected, under which only two of the Weyl fermions and two of the complex scalars are charged.},
the absence of truly non-abelian structures in the actual computation makes the effect of the sum over all the 15
generators $T^I$ in the vertex with the ``$SU(4)_R$ photons'' practically equivalent to the consideration of 15 $U(1)$ photons.
 
Together with the presence of many more processes due to scalar contributions both in virtual and real diagrams, the main difference with respect to the QCD setting stays in the necessity of assigning appropriate R-symmetry (flavor) factors to the squared diagrams and interference terms. This is due to the presence, in the Yukawa-type vertices of $\N=4$ SYM lagrangian, of sigma-matrices connecting the vectorial and spinorial representations of $SO(6)_R\sim SU(4)_R$, under which the $\N=4$ SYM scalar and fermionic fields, respectively, rotate. In performing the calculation, we have used Feynman rules derived from the $\N=4$ SYM lagrangian in the notation of~\cite{Beisert:2004ry} and the FDH regularization scheme of~\cite{Bern:1991aq,Bern:2002zk}.






 The ``DIS'' Wilson coefficients of  $\mathcal{N}=4$ SYM are then given at NLO (order $\hat a=\frac{\alpha N_c}{4\pi}$) by the following formulas, extremely simple and symmetric in scalar and fermionic contributions,  
\begin{align}\label{result1}
  C_\la(z)&=2 C_A \Big[\hat P^{(0)}_\la(z)\, \ln\big(\frac{z}{1-z}\big)+ 2\zeta_2 C_{(f)}\delta(1-z)\Big]\,,\\\label{result2}
  C_g(z)&=2 C_A \Big[\hat P^{(0)}_g(z) \,\ln\big(\frac{z}{1-z}\big)\,\Big]\,,\\\label{result3}
  C_\phi(z)&=2 C_A \Big[\hat P^{(0)}_\phi(z) \,\ln\big(\frac{z}{1-z}\big)+2\zeta_2 C_{(v)}\delta(1-z)\Big]\,,
  \end{align}
where $\hat P$ is defined as the following linear combination of splitting functions
\be\label{Phat}
\hat P^{(0)}_{a}(z)=P^{(0)}_{\la a}(z)C_{(f)}+  P^{(0)}_{\phi a}(z) C_{(v)}\quad \quad a=g,\phi,\la.
\ee
Above, the LO splitting functions are those reported in Appendix \ref{app:splittingfunctions}, formulas (\ref{splitN=4first})-(\ref{splitN=4last}),
 $C_{(v)}$ and $C_{(f)}$ are the quadratic Casimir invariants respectively in the vectorial representation of SO(6) and fundamental of $SU(4)$ (the latter is equivalent to the spinorial representation of SO(6)), $C_A$ is the Casimir eigenvalue for the adjoint representation of the gauge group $SU(N_c)$.
Already in this form, it is not difficult to see that the first two functions $C_\lambda$ and $C_g$ above, with the appropriate truncation of the scalar sector, coincide with the highest transcendental part of their QCD counterpart (\ref{CqQCD})-(\ref{CgQCD})~\footnote{For the definition of the degree of transcendentality for a general function, see for example~\cite{Kotikov:2013gga,Henn:2013pwa} and references therein.}.
In Mellin space, formulas (\ref{result1})-(\ref{result3}) read
\begin{align}
  C_\la(j)&=2 C_{(f)} C_A \, \Big[ 2\,S_2(j)+4\frac{S_1(j)}{j}-2S_{11}(j)-\frac{4}{j^2}\Big]\,, \label{resmel1}\\
  C_g(j)&=2 C_{(f)} C_A \, \Big[4\,\frac{S_1(j)}{j}-\frac{4}{j^2}\Big]\,, \label{resmel2}\\
  C_\phi(j)&=2 C_{(v)} C_A \,\Big[2S_2(j)+3\frac{S_1(j)}{j}-2S_{11}(j) -\frac{3}{j^2}\Big]~.\label{resmel3}
  \end{align}
Assigning to  $\frac{1}{j^n}$ transcendentality degree $n$, the expressions above exhibit uniform transcendentality of degree $2$, which is the maximum degree expected at this order in perturbation theory\footnote{It is well-known~\cite{Magnea:1990zb}  that the maximum order of soft and collinear divergences of the Sudakov form factor at one loop is $\frac{1}{\epsilon^2}$.}. In that formulas above are \emph{not only} given in terms of harmonic sums of fixed degree 2, their uniform, leading transcendentality feature is sometimes referred to as ``weak'', as opposed to the one in ``strong'' sense which is the argument of next section.

\section{Change of basis and prediction at NNLO}
\label{sec:diagonalization}
\setcounter {equation} {0}

Experience in the calculation of anomalous dimensions~\cite{Kotikov:2002ab,Kotikov:2004er} of leading twist operators in $\N=4$ SYM suggests that the $n$-loop anomalous dimension matrix $\gamma_{ab}(j)$ ($a,b=g,\lambda,\phi$), after diagonalization, assumes an intriguing form in terms of a single universal function $\ga_{uni}(j)$ expressed only through a combination of harmonic sums of constant degree $2n-1$. Explicit expressions for the matrices entering the diagonalization can be found in Appendix~\ref{app:diagonalization}. It is quite natural to ask how Wilson coefficients would appear in this new basis. 

 Let us first notice that the particular combination of splitting functions \eqref{Phat} appearing in formulas \eqref{result1}-\eqref{result3} would also appear as coefficient of $\log\big(\frac{Q^2}{\mu^2}\big)$ in the $\N=4$ analogous of \eqref{Fq}-\eqref{Fg}. In Mellin space it is therefore natural to define the vector 
\be\label{vectorandim}
\bar{\gamma}(j)=(\bar{\gamma}_g(j),\bar{\gamma}_\la(j),\bar{\gamma}_\phi(j)),
\ee
with\footnote{The coefficient $\frac43$ comes from the ratio $\frac{C_{(v)}}{C_{(f)}}$ obtained after collecting $C_{(f)}$ in the Mellin transform of (\ref{Phat}).} $\bar{\gamma}_a(j)=\ga^{(0)}_{\la a}(j)+\frac{4}{3}\ga^{(0)}_{\phi a}(j)$, $a=g,\la,\phi$, and rotate it using the same matrix (\ref{Vinv}) used in~ \cite{Kotikov:2002ab,Kotikov:2004er}. Setting $v_g=2\frac{1-j}{1-2j}$ and $v_\la=\frac{1}{1-2j}$ in \eqref{Vmat} and \eqref{Vinv}, the result is
\be\label{divrot}
\hat{\gamma}(j)=V^{-1}\bar{\gamma}(j)=\left(\ga^{(0)}_{uni}(j-2),\ga_{uni}^{(0)}(j),0\right),
\ee
where $\ga_{uni}^{(0)}(x)=-4 S_1(x)$. 
Turning to the Wilson coefficients it is natural to define
\be
C(j)=(C_g(j),C_\la(j),C_\phi(j)),
\ee
where the three components are given in (\ref{resmel1})-(\ref{resmel3})\footnote{From now on an overall factor $C_{(f)} C_A$ will be neglected.}. Rotating again with $V^{-1}$ given in \eqref{Vinv} one gets
\be\label{resrot}
\hat{C}(j)=V^{-1}C(j)=\left(C_{uni}(j-2)-\frac{4}{(j-1)^2},C_{uni}(j),0\right)
\ee
where we defined
\be\label{resuniv}
C_{uni}(j)=4(S_{1,1}(j)-S_2(j)).
\ee
Comparing this result with the QCD one-loop coefficient functions in Mellin space \cite{Moch:1999eb}, one realizes that the maximally transcendental part of that result is given precisely by the combination (\ref{resuniv}) of harmonic sums of degree 2.  It is therefore for the Wilson coefficients \emph{rotated vector} that the proper maximum transcendentality principle works,  in total analogy with \cite{Kotikov:2002ab}. 
For the case presented here, however, a couple of new interesting features appear.
The main difference with respect to the structure of the diagonal expression (\ref{Vdiag}) is the vanishing of the  third component for the vectors (\ref{divrot}) and (\ref{resrot}). It is quite interesting that the same feature is present in both cases and it would be interesting to check it at NNLO.
The other novel feature is the shift of $\frac4{(j-1)^2}$ in the first component (\ref{resrot}). It is worthwhile to notice that this additional term is responsible for the cancellation of the singularity appearing in $ C_{uni}(j-2)$ in the Regge limit $j\to1$. 

To better understand the pole structure in the Regge limit, it is interesting to consider the QCD case~\cite{vanNeerven:1991gh,Moch:1999eb,Vermaseren:2005qc}, where the 
gluon Wilson coefficient $\hat{F}_g^{\rm QCD}(1,j)$ has the following form
\be
\hat{F}_{g}^{\rm QCD}(1,j)=1 + \sum_{n=1}^{\infty} {\left(\frac{\alpha_s(Q^2)}{4\pi}\right)}^n  C_g^{(n),\rm QCD} (j) \, ,
\label{WCoeQCD}
\ee
which, for $j\to 1$, reads
\begin{eqnarray}
&&  C_g^{(1),\rm QCD} (j) = \frac{2}{3} n_f + \mathcal{O}((j-1)^1)\, , \nonumber  \\
&&  C_g^{(2),\rm QCD} (j) = \frac{8}{3} n_f C_A \left[\frac{43}{9}-2\zeta_2 \right] \frac{1}{j-1} + \mathcal{O}((j-1)^0)\, , \nonumber  \\
&&  C_g^{(3),\rm QCD} (j) = \frac{32}{9} n_f C_A^2 \left[\frac{1234}{27}-13\zeta_2+4\zeta_3 \right] \frac{1}{(j-1)^2}
+ \mathcal{O}((j-1)^{-1})\, , 
\label{WCoeQCD1}
\end{eqnarray}
with $ n_f$ the number of active quarks in QCD.
Starting from NNLO ($n=2$) above, the most singular terms in $ C_g^{(n),\rm QCD} (j)$ contain $(j-1)^{1-n}$ contributions and are proportional to the factor $ n_f$.
 Direct inspection of the QCD results shows that, both for anomalous dimensions and coefficient functions, the contributions with a factor $ n_f$ do not contain terms with maximal transcendentality, $2n$, expected at a given loop-order $n$.  So, according to the MTP, the terms  $\sim n_f$, including the singular ones in (\ref{WCoeQCD1}), should not contribute to the $\N=4$ SYM case.
Accepting, by analogy with the case of anomalous dimension matrix, that at all orders in perturbation theory there
is one universal Wilson coefficient function obeying the MTP, it is easy to see that the only possible poles in its '$+$' component~\footnote{In analogy with the anomalous dimension case, it is the first component of the rotated Wilson coefficient vector, in which $C_{uni}(j-2)$ appears, which exhibits singularities in the Regge limit.}  -- i.e. the first one of the vector in (\ref{resrot}) -- are combinations of
 terms $1/(j-1)^{2n}$ and 
$\zeta_m/(j-1)^{2n-m}$ with $2\leq m \leq 2n-1$. Since such terms are absent in the QCD case, and $ C^{(n)}_{uni}(j)$  should contain only
 the most complicated terms of $C_g^{(n),\rm QCD} (j)$, we conclude that no singular terms should appear in $ C^{(n)}_{+}(j)$ (see eqs. \eqref{+1a} and  \eqref{+2a} below). 
 This would imply that all the singularities for $ j\to 1$ are contained in the first component of the anomalous dimension matrix (or, equivalently, in the first component of vector \eqref{divrot}). 

Another argument supporting the finiteness of the coefficient function in the Regge limit comes from the analysis of BFKL equation for $\mathcal{N}=4$ SYM. The correspondence between the singularities for $j\to 1$ obtained by the BFKL approach and those appearing in the ``+'' component of the anomalous dimension (i.e. $\gamma_{uni} (j-2)$) has been checked up to NLO and it is believed to hold at all orders in perturbation theory \cite{Kotikov:2002ab}. In other words, the ``$+$'' component of the diagonalized anomalous dimensions matrix in QCD and in $\mathcal{N}=4$ SYM can be reconstructed from the BFKL approach in the limit $j \to 1$ (see \cite{Kotikov:2002ab,Fadin:1998py,Kotikov:2000pm})\footnote{Several authors speak about the reconstruction of gluon anomalous dimension, but this is not true beyond LO. What is reconstructed is the first component of the diagonalized matrix (see the recent discussion in \cite{Bolzoni:2012ii}).}. Hence, assuming that the BFKL equation collects all singular terms, for $j\
to1$, of the $\mathcal{N}
=4$ SYM structure 
function $\hat{F}_{uni} (Q^2/\mu^2 , j)$, all the singularities should appear only in the anomalous dimension $\gamma_{uni} (j-2)$ and not in the Wilson coefficients.


All these arguments suggest that the ``+'' component of the Wilson coefficient rotated vector should be finite in the Regge limit.  It would be interesting to check this conjecture at NNLO by explicit computation.

Following these considerations we can guess the form of $C_{uni}(j)$ at NNLO from the corresponding QCD result~\cite{Moch:1999eb}. The structure of the result in Mellin space, in analogy with (\ref{F2QCD}), should be
\begin{eqnarray}
&&\hat{F}_{uni}\Big(\frac{Q^2}{\mu^2},j\Big)=1 + \hat a \Bigl[C^{(1)}_{uni}(j) + \gamma^{(0)}_{uni}(j) 
\ln\Big(\frac{Q^2}{\mu^2}\Big) \Bigr]\\ \nonumber
&&+~ \hat{a}^2 \Bigl[C^{(2)}_{uni}(j)+\left(\gamma^{(1)}_{uni}(j) + \gamma^{(0)}_{uni}(j) C^{(1)}_{uni}(j) \right) \ln\Big(\frac{Q^2}{\mu^2}\Big) + \frac{1}{2} {\left(\gamma^{(0)}_{uni}(j)\right)}^2 \ln^2\Big(\frac{Q^2}{\mu^2}\Big) \Bigr]+\mathcal{O}(\hat{a}^3)\, ,
\end{eqnarray}
where~\footnote{Note that, as in Ref. \cite{Fadin:1998py,Ciafaloni:1998gs}, our normalization of $\gamma(j)$ contains an extra factor $-1/2$ with respect to standard literature (see \cite{LONLOAD}) and differs by a sign in comparison with \cite{Moch:2004pa}.}
\begin{eqnarray}
\frac{1}{2} \, C^{(1)}_{uni} &=& S_1^2 -  S_2,  \label{Cuni1} \\
\frac{1}{2} \, C^{(2)}_{uni} &=& 
S_1^4 - 8\,S_1^2 \Bigl(S_{2} +  \overline S_{-2} \Bigr) + 4\,S_1
\Bigl(\overline S_{-3} +  2\,\overline S_{-2,1} \Bigr) 
+ S_2^2 + 4\,S_2\,\overline S_{-2} +
6\,\overline S_{-2}^2\nonumber \\
&& + 10\,\overline S_{-4} + 2\,S_{4} + 8\,S_{3,1}
- 8\,\overline S_{-3,1} - 4\,\overline S_{-2,2} \label{Cuni2}\ .
\end{eqnarray}
The definition of the nested harmonic sums $S_{\pm a,\pm b,\pm c,\cdots}(j)$ is given in Appendix \ref{app:harmonicsums}. The extraction of $C_{uni}^{(3)}$ from the QCD result~\cite{Vermaseren:2005qc} is rather complicated, and here we will report only the expression for $j\to \infty$. In this limit and up to three loops, one finds~\footnote{We are grateful to G. Korchemsky for pointing out a misprint in the three-loop result in the first version of the paper. }
\begin{eqnarray}
C^{(1)}_{uni}(j) &=& 2\ln^2 \frac{j}{j_0} -2\, \zeta_2+O(j^{-1}),  \label{jlarge.1} \\
C^{(2)}_{uni}(j) &=& 2\ln^4 \frac{j}{j_0} -8\, \zeta_2\ln^2 \frac{j}{j_0} -16\, \zeta_3\ln \frac{j}{j_0} +\frac{55}{2}\, \zeta_4+O(j^{-1}), 
 \label{jlarge.2} \\
\!\!\!\!\!\!\!\!\!\!\!\!\!\!\!\!\!\!
C^{(3)}_{uni}(j) &=& \frac{4}{3}\ln^6 \frac{j}{j_0} -12\, \zeta_2\ln^4 \frac{j}{j_0} -
32\, \zeta_3\ln^3 \frac{j}{j_0} + 119\,  \zeta_4\ln^2 \frac{j}{j_0} \nonumber\\
&&-\big(352\,\zeta_5 - \textstyle{\frac{224}{3}}\,\zeta_2\zeta_3\big) \ln \frac{j}{j_0} + \frac{112}{3}\, {\zeta_3}^2-\frac{13485}{72}\, \zeta_6+O(j^{-1})\, , ~~~~~~
\label{jlarge.3}
\end{eqnarray}
with $j_0=e^{-\gamma_E}$.

Let us comment on the agreement of expressions \eqref{jlarge.1}-\eqref{jlarge.3} with the expected results coming from exponentiation of the large j logarithms~\footnote{See also~\cite{Alday:2013cwa} for  a prediction on the large $j$  behavior of the universal structure constants of $\mathcal{N}=4$ SYM, which can be compared - invoking the OPE analysis of DIS - with our predictions \eqref{jlarge.1}-\eqref{jlarge.3}. }. From general arguments \cite{Catani:1989ne,Sterman:1986aj,Korchemsky:1993uz} it is known that in DIS, for a general massless gauge theory, large logarithms for $j\to\infty$ are controlled by the cusp anomalous dimension $\Gamma_{\textup{cusp}}(\hat a)$ and an additional anomalous dimension $\Gamma_{\textup{DIS}}(\hat a)$. In the case of $\N=4$ SYM, with vanishing $\beta$-function, including finite contributions for $j\to\infty$ along the lines of \cite{Vogt:1999xa} one gets
\be\label{resum}
C_{uni}(j)\sim g(\hat a) \,e^{G\left(\hat a,\frac{j}{j_0}\right)},
\ee
where $g(\hat a)$ collects all the finite contributions of $C_{uni}(j)$ for $j\to\infty$. For the simple case of $\N=4$ SYM the function $G\left(\hat a,\frac{j}{j_0}\right)$ can be written as
\be
G\left(\hat a,\frac{j}{j_0}\right)=\frac{1}{2}\,\Gamma_{\textup{cusp}}(\hat a) \ln^2\frac{j}{j_0}+\Gamma_{DIS}(\hat a) \ln\frac{j}{j_0}~.
\ee
The anomalous dimensions and the function $g(\hat a)$ clearly admit a perturbative expansion in powers of $\hat{a}$. From our result \eqref{jlarge.1}-\eqref{jlarge.3} we can read out the first terms of those expansions,  getting
\begin{align}
g(\hat a)&=1-2\,\hat a \zeta_2+\frac{55}{2} \hat a^2 \zeta_4 +\hat a^3\left(\frac{112}{3} {\zeta_3}^2-\frac{13485}{72} \zeta_6\right)+O(\hat a^4)\,,\\
\Gamma_{\textup{cusp}}(\hat a)&=4\, \hat a-8\,\hat a^2 \zeta_2+88\, \hat a^3\zeta_4+O(\hat a^4)\,,\\
\Gamma_{DIS}(\hat a)&=-16\, \hat{a}^2 \zeta_3+ 32\hat a^3 \left(\frac43 \zeta_2 \zeta_3 + 11 \zeta_5\right)+O(\hat a^4)~.
\end{align}
It is noteworthy that the function $\Gamma_{\textup{cusp}}(\hat a)$ coincides up to three loops with the well-known expansion of the cusp anomalous dimension of $\N=4$ SYM \footnote{Our convention for the cusp anomalous dimension follows the one used in \cite{Beisert:2006ez}. The function $\gamma_k(a)$ defined in \cite{Bern:2006ew} is given by $\gamma_k(a)=2 \Gamma_{\textup{cusp}}(\hat a)$ with $a=2 \hat a$.} \cite{Beisert:2006ez,Bern:2006ew}. This provides a good compatibility check for our result.

Let us now consider the limit of $C^{(2)}_{uni}(j)$ for $j\to -1$\footnote{This is equivalent to study the limit $j\to1$ of $C^{(2)}_{uni}(j-2)$, which is the function appearing as the first component of \eqref{resrot}.}. To this purpose it is useful to recast the coefficient functions \eqref{Cuni1}-\eqref{Cuni2} in the form~\footnote{As $j\to -1$, nested harmonic sums $S_{\pm a,\pm b, ...}(j)$ have a less singular behaviour than standard sums $S_{\pm a}(j)$~\cite{Kotikov:2004er}. }
\begin{eqnarray}
C^{(1)}_{uni}(j) &=& 4\,\Bigl(S_{1,1}- S_2\Bigr),  \label{Cuni1a} \\
C^{(2)}_{uni}(j) &=& 48\,S_{1,1,1,1} - 56\,\Bigl(S_{2,1,1}+S_{1,2,1}+S_{1,1,2} \Bigr)
+ 56\,S_{3,1} +48\, S_{2,2} + 24\,\Bigl(\overline S_{2,-2}+\overline S_{-2,-2}\Bigr)  \nonumber \\
&& - 16\,\Bigl(\overline S_{1,-2,1}+2 \,\overline S_{1,1,-2}\Bigr) + 8\,\overline S_{-3,1}
+40\,\Bigl(S_{1,3}+\overline S_{1,-3}\Bigr) - 12\,\overline S_{-4}- 28\,S_4,
\label{ha2a}\label{Cuni2a}
\end{eqnarray} 
The asymptotic behavior of the expressions above can be extracted from Appendix A of \cite{Kotikov:2004er}, putting $r=-1$.
The assumption of finiteness as $j\to1$ for the ``+'' component of the vector $\hat{C}$  leads to the following conjecture for its explicit form at NNLO
\begin{eqnarray}
C^{(1)}_{+}(j) &=& C^{(1)}_{uni}(j-2) - \frac{4}{(j-1)^2},  \label{+1a} \\
C^{(2)}_{+}(j) &=& C^{(2)}_{uni}(j-2) - \frac{16}{(j-1)^4}
+ \frac{48}{(j-1)^2}\zeta_2 -  \frac{40}{j-1}\zeta_3 \label{+2a}~.
\end{eqnarray}
It would be interesting to verify this conjecture by explicit computation. 

\section*{Acknowledgments}

We thank L.J.~Dixon, G.~Korchemsky, L.~Lipatov, L.~Magnea, C.~Meneghelli, J.~Plefka and V.~Schomerus for useful discussions, and G.~Korchemsky for noticing a misprint  in the first version of the paper.
V.F. is particularly grateful to L.J.~Dixon for suggesting the study of generalized cross-sections in $\mathcal{N}=4$ SYM and for earlier collaboration on related topics. The work of A.V.K. was supported in part by the Russian Foundation for Basic Research
(Grant No. 13-02-01005). The work of L.B. and V.F. is funded by the the German Research Foundation (DFG) via the Emmy Noether Program ``Gauge Fields from Strings''. 

\appendix

\section{Splitting functions in QCD and in $\N=4$ SYM}
\label{app:splittingfunctions}

\def\theequation{A.\arabic{equation}}
\setcounter{equation}{0}

The leading order contributions to the DGLAP splitting functions in QCD read
\begin{eqnarray}\label{Pqq}
P_{qq}^{(0)}(x)&=&C_F\,\{2\,p_{qq}(x)+3\,\delta(1-x)\,\}\,,\\
P_{qg}^{(0)}(x)&=&2\,n_f\,p_{qg}(x)\,,\qquad\qquad
P_{gq}^{(0)}(x)=2 C_F\,p_{gq}(x)\,,\\
P_{gg}^{(0)}(x)&=&4 C_A\,p_{gg}(x)+\frac{11\,C_a+4\,n_f\,T_R}{6}\delta(1-x)
\end{eqnarray} 
with $T_R=\frac{1}{2}$ and  
\begin{eqnarray}\label{psmallqq}
p_{qq}(x)&=&\frac{2}{1-x}-1-x\,,\qquad\qquad
p_{qg}(x)=1-2x+2x^2\,,\\\label{psmallgg}
p_{gq}(x)&=&\frac{2}{x}-2+x\,,\qquad\qquad
~~~~~p_{gg}(x)=\frac{1}{1-x}+\frac{1}{x}-2+x+x^2.
\end{eqnarray}
Their corresponding, Mellin-transformed, anomalous dimensions  are
\ba\label{gammaQCD}
\gamma_{gg}^{(0)}(j)  &=& 
2C_A\bigg[ - S_1(j) +\frac{1}{j(j-1)}+\frac{1}{(j+1)(j+2)}
           + \frac{11}{12} \bigg] - \frac{2}{3} n_f\,T_R \,,\\
\gamma_{gq}^{(0)}(j) &=& C_F\frac{j^2+j+2}{j(j^2-1)}\,,\qquad\qquad 
\gamma_{qg}^{(0)}(j) = T_R\frac{j^2+j+2}{j(j+1)(j+2)} \,,\\
\gamma_{qq}^{(0)}(j) &=& 
C_F \bigg[ - 2 S_1(j) + \frac{1}{j(j+1)} + \frac{3}{2} \bigg] \,.
\ea
The splitting functions for  twist-two operators in $\N=4$ SYM 
are given by~\cite{Kotikov:2002ab}
\begin{align}\label{splitN=4first}
  P^{(0)}_{\la\la}(z)&=C_A \left(\frac{1+z^2}{(1-z)_+}+3(1-z)\right); &P^{(0)}_{\phi\la}(z)&=C_A 3z;\\
   P^{(0)}_{\la g}(z)&=C_A \,4\left(1-2z+2z^2\right); &P^{(0)}_{\phi g}(z)&= C_A 6(z-z^2);\\\label{splitN=4last}
  P^{(0)}_{\la\phi}(z)&=C_A \,4; &P^{(0)}_{\phi\phi}(z)&=C_A \frac{2z}{(1-z)_+}.
\end{align}
Their corresponding anomalous dimensions are 
\ba\label{gammaN4}
\g_{gg}^{(0)}(j) &=&    2\, C_A\,\Big[-S_1(j)+\frac{1}{j(j-1)}+\frac{1}{(j+1)(j+2)}\Big]\\
\g_{g \la}^{(0)}(j) &=&   C_A\,\frac{(j^2+j+2)}{j(j^2-1)},  ~~~~~~~~~~~~~~~~~~~~~~~~
\g_{\phi g}^{(0)}(j) =6 \,C_A\,\Big[\frac{1}{j+1}-\frac{1}{j+2}\Big]\\
\g_{\la\la}^{(0)}(j) &=& 2\, C_A\,\Big[-S_1(j)+\frac{2}{j(j+1)} \Big],   ~~~~~~~~~~
\g_{\la\phi}^{(0)}(j) =C_A\,\frac{4}{j} \\
 \g_{\la g}^{(0)}(j) &=& 4\,C_A\, \frac{j^2+j+2}{j(j+1)(j+2)}, ~~~~~~~~~~~~~~~~~~
\g_{\phi\la}^{(0)}(j) =C_A\,\frac{3}{j+1} \\
\g_{\phi\phi}^{(0)}(j) &=& -2\,C_A\, \,S_1(j), ~~~~~~~~~~~~~~~~~~~~~~~~~~~~~
\g_{g\phi}^{(0)}(j) =2\,C_A\,\Big[\frac{1}{j-1} -\frac{1}{j} \Big]
\ea


\section{Diagonalization of anomalous dimensions in $\mathcal{N}=4$ SYM}
\label{app:diagonalization}

\def\theequation{B.\arabic{equation}}
\setcounter{equation}{0}
In~\cite{Kotikov:2002ab} it was seen that the matrix $\gamma_{ab}(j)$ ($a,b=g,\lambda,\phi$) is diagonalized by
\be\label{Vmat}
V=\left(\begin{array}{ccc}
v_g&-2(j-1) v_q & \frac{j(j-1)}{(j+1)(j+2)} v_\phi\\
v_g& v_q &\frac{j}{j+1}v_\phi \\
v_g & \frac23 (j+1)v_q &v_\phi\end{array} \right). 
\ee
Namely
\be\label{Vdiag}
V^{-1} \ga V=\left(\begin{array}{ccc}
\ga_{uni}(j-2)&0 &0\\
0& \ga_{uni}(j) &0 \\
0 & 0 &\ga_{uni}(j+2)\end{array} \right),
\ee
with
\footnote{The results for $V^{-1}$ in \cite{Kotikov:2002ab} have two slips.
The intersection of the second column and the third row should contain
an additional minus sign. The intersection of the  third column and the
second row should contain an additional factor 3.}
\be\label{Vinv}
V^{-1}= \frac{(j+1)(j+2)}{2(4j^2-1)(2j+3)} \, \left(\begin{array}{ccc}
(2j+3)v_g^{-1}&4\frac{j-1}{j+2}(2j+3)v_g^{-1} & 
3\frac{j(j-1)}{(j+1)(j+2)}(2j+3)v_g^{-1}\\
-3\frac{2j+1}{j+1}v_q^{-1}& 6\frac{2j+1}{(j+1)(j+2)}v_q^{-1} &
3\frac{j(2j+1)}{(j+1)(j+2)}v_q^{-1} \\
(2j-1)v_\phi^{-1} & -4(2j-1)v_\phi^{-1} &3(2j-1)v_\phi^{-1} \end{array}\right).
\ee
The explicit expression of $\ga_{uni}$ up to NNLO 
is given in \cite{Kotikov:2004er}.

\section{Harmonic sums}\label{app:harmonicsums}
\def\theequation{C.\arabic{equation}}
\setcounter{equation}{0}

The basic definition of standard and nested harmonic sums with general indices are 
\be  \label{ha1} S_{\pm a}(j) = \sum^j_{m=1} \frac{(\pm 1)^m}{m^a},~~
S_{\pm a,\pm b,\pm c,\cdots}(j) = \sum^j_{m=1} \frac{(\pm 1)^m}{m^a} S_{\pm b,\pm c,\cdots}(m), 
\ee
where we omit the sign ``$+$''.
The nested harmonic sums appearing in (\ref{Cuni2a}) are defined as
\begin{eqnarray} \label{ha3}
&&\!\!\!\!\!\!\!\!\!\!\!\!
\overline S_{-a,b,c,\cdots}(j) = (-1)^j \, S_{-a,b,c,...}(j)+ S_{-a,b,c,\cdots}(\infty) \, \Bigl( 1-(-1)^j \Bigr), \\\nonumber
&&\!\!\!\!\!\!\!\!\!\!\!\!
\overline S_{-a,-b,c,\cdots}(j)= S_{-a,-b,c,...}(j)
+  \Bigl( 1-(-1)^j \Bigr) S_{-b,c,\cdots}(\infty) \Bigl[S_{-a}(\infty) - S_{-a}(j) \Bigr], \\\nonumber
&&\!\!\!\!\!\!\!\!\!\!\!\!
\overline S_{a,-b,c,\cdots}(j)= (-1)^j \, S_{a,-b,c,...}(j) +  \Bigl( 1-(-1)^j \Bigr) \Bigl(S_{a,-b,c,\cdots}(\infty) - S_{-b,c,\cdots}(\infty) 
\Bigl[S_{a}(\infty) - S_{a}(j) \Bigr]\Bigr), \\\nonumber
&&\!\!\!\!\!\!\!\!\!\!\!\!
\overline S_{a,b,-c,\cdots}(j) = (-1)^j \, S_{a,b,-c,...}(j)
+  \Bigl( 1-(-1)^j \Bigr) \Bigl(S_{a,b,-c,\cdots}(\infty)-S_{-c,\cdots}(\infty) \Bigl[S_{a,b}(\infty) - S_{a,b}(j) \Bigr]
\nonumber \\ 
&&\qquad\qquad~
- \Bigl(S_{b,-c,\cdots}(\infty)
- S_{b}(\infty)S_{-c,\cdots}(\infty)\Bigr)\Bigl[S_{a}(\infty) - S_{a}(j) \Bigr]
\Bigr).
\label{ha3a}
\end{eqnarray}
Expressions (\ref{ha3})- (\ref{ha3a}) are defined for integer values of the arguments (see \cite{Kazakov:1992xj,Kotikov:2001sc}), but can be easily analytically continued to real and complex $j$ via the methods of Refs.~\cite{Kotikov:2002ab,Kotikov:1994mi,Kotikov:2001sc,Kotikov:2005gr}.

\AtEndEnvironment{thebibliography}{
\bibitem{reciprocity} 
 M.~Beccaria and V.~Forini,
  JHEP {\bf 0806}, 077 (2008)
  [arXiv:0803.3768 [hep-th]];
   V.~Forini and M.~Beccaria,
  Theor.\ Math.\ Phys.\  {\bf 159}, 712 (2009)
  [Teor.\ Mat.\ Fiz.\  {\bf 159}, 353 (2009)]
  [arXiv:0810.0101 [hep-th]];
  M.~Beccaria and V.~Forini,
  JHEP {\bf 0903}, 111 (2009)
  [arXiv:0901.1256 [hep-th]];
   M.~Beccaria, V.~Forini and G.~Macorini,
  Adv.\ High Energy Phys.\  {\bf 2010}, 753248 (2010)
  [arXiv:1002.2363 [hep-th]].
\bibitem{LONLOAD}  D.~J.~Gross and F.~Wilczek, Phys.\ Rev.\ D \textbf{8}
(1973) 3633; H.~Georgi and H.~D.~Politzer, Phys.\ Rev.\ D \textbf{9} (1974)
416; E.~G.~Floratos, D.~A.~Ross and C.~T.~Sachrajda, Nucl.\ Phys.\
\textbf{B129} (1977) 66; [Erratum-ibid.\ \textbf{B139} (1978) 545]; E.
G.~Floratos, D.~A.~Ross and C.~T.~Sachrajda, Nucl.\ Phys.\ \textbf{B152}
(1979) 493; A.~Gonzalez-Arroyo, C.~Lopez and F.~J.~Yndurain, Nucl.\ Phys.\
\textbf{B153} (1979) 161; A.~Gonzalez-Arroyo and C.~Lopez, Nucl.\ Phys.\
\textbf{B166} (1980) 429;
G.~Gurci, W.~Furmanski and R.~Petronzio, Nucl.\ Phys.\
\textbf{B175} (1980) 27; W.~Furmanski and R.~Petronzio, Phys.\ Lett.\
\textbf{B97} (1980) 437; E.~G.~Floratos, C.~Kounnas and R.~Lacage, Nucl.\ Phys.\
\textbf{B192} (1981) 417; C.~Lopes and F.~J.~Yndurain, Nucl.\ Phys.\
\textbf{B171} (1980) 231, \textbf{B183} (1981) 157;
R.~Hamberg and W.~L.~van Neerven, Nucl.\ Phys.\ \textbf{B379} (1992) 143;
R.~K.~Ellis and W.~Vogelsang, arXiv:hep-ph/9602356;
R.~Mertig and W.~L.~van Neerven, Z.\ Phys.\ \textbf{C70}
(1996) 637; W.~Vogelsang, Nucl.\ Phys.\ \textbf{B475} (1996) 47.
}

\bibliographystyle{nb}

\bibliography{refWC}

\end{document}